\documentclass[11pt]{article}

\usepackage[top=1in, left=1in, right=1in,
bottom=1in]{geometry}

\usepackage{latexsym}
\usepackage[parfill]{parskip}
\usepackage{amssymb,amsmath, bm}
\usepackage{graphicx}
\usepackage{marvosym}
\usepackage{lscape}
\usepackage{rotating}
\usepackage{setspace}
\usepackage{threeparttable}
\usepackage{booktabs}
\usepackage[compact]{titlesec}
\usepackage{multirow}

\usepackage{caption}
\usepackage{subcaption}

\usepackage{bbm}

\usepackage{lmodern}
\fontfamily{lmtt}\selectfont
\usepackage[T1]{fontenc}

\titleformat{\section}
  {\normalfont\fontsize{11}{15}\bfseries}{\thesection}{1em}{}

\titleformat{\subsection}
  {\normalfont\fontsize{11}{15}\bfseries}{\thesubsection}{1em}{}

\usepackage{natbib}
\usepackage{color}
\usepackage[pdftex, bookmarksopen=true, bookmarksnumbered=true,
pdfstartview=FitH, breaklinks=true, urlbordercolor={0 1 0}, citebordercolor={0 0 1}]{hyperref}

\usepackage{dcolumn}
\newcolumntype{.}{D{.}{.}{-1}}
\newcolumntype{d}[1]{D{.}{.}{#1}}

\usepackage{amsthm}

\def\b1{\boldsymbol{1}}

\begin{document}

\pagestyle{plain}

\newcommand{\blind}{0}

\newcommand{\tit}{\Large Using Balancing Weights to Target the Treatment Effect on the Treated when Overlap is Poor}

\if0\blind

{\title{\tit}
\author{Eli Ben-Michael\thanks{Carnegie Mellon University, Email: ebenmichael@cmu.edu}
\and Luke Keele\thanks{University of Pennsylvania, Philadelphia, PA, Email: luke.keele@gmail.com}
}

\date{\today}

\maketitle
}\fi

\if1\blind
\title{\bf \tit}
\maketitle
\fi

\begin{abstract}

Inverse probability weights are commonly used in epidemiology to estimate causal effects in observational studies. Researchers can typically focus on either the average treatment effect or the average treatment effect on the treated with inverse probability weighting estimators. However, when overlap between the treated and control groups is poor, this can produce extreme weights that can result in biased estimates and large variances. One alternative to inverse probability weights are overlap weights, which target the population with the most overlap on observed characteristics. While estimates based on overlap weights produce less bias in such contexts, the causal estimand can be difficult to interpret. One alternative to inverse probability weights are balancing weights, which directly target imbalances during the estimation process. Here, we explore whether balancing weights allow analysts to target the average treatment effect on the treated in cases where inverse probability weights are biased due to poor overlap. We conduct three simulation studies and an empirical application. We find that in many cases, balancing weights allow the analyst to still target the average treatment effect on the treated even when overlap is poor. We show that while overlap weights remain a key tool for estimating causal effects, more familiar estimands can be targeted by using balancing weights instead of inverse probability weights. \\

\noindent causal inference; weighting, causal estimands, inverse probability weighting, balancing weights, overlap weights

\end{abstract}

\clearpage
\doublespacing

\section{Introduction}

Evidence for the causal effect of an intervention is often drawn from observational studies, especially in settings where randomized trials are infeasible. The first step of conducting an observational study is defining the causal effect---i.e. estimand---of interest. Two common estimands targeted in observational studies are the average treatment effect (ATE) and the average treatment effect on the treated (ATT). The ATE measures the average difference in outcomes when all individuals in the study population are assigned to treatment versus when all individuals are assigned to control. The ATT, on the other hand, measures the average difference in outcomes among those individuals in the population that were actually exposed to the treatment. These estimands answer different scientific questions, so investigators must select which to target based on substantive judgements.

Selection of the estimand may also depend on estimation considerations. Methods such as matching and inverse probability weighting (IPW) can be used to target either the ATE or the ATT. However, both methods depend on the assumption that each unit in the study has a non-zero probability of being treated.  Violation or near violation of this assumption can result in biased or imprecise estimates \citep{Crump:2009}. Critically, when overlap between the treated and control populations is limited, the ATT may be identifiable when the ATE is not, for instance when only some units have a non-zero probability of treatment. For some data configurations, overlap may be so limited that even the ATT may not be identifiable. When this occurs, one strategy is to use an alternative estimand that only targets the subset of treated units that overlap with the control units \citep{Crump:2009,Rosenbaum:2011,li2018balancing}. One such estimand is the average treatment effect for the overlap population (ATO) \citep{li2018balancing}. Under the ATO, the estimand is focused on the  marginal population that 0might or might not receive the treatment of interest rather than a known, a priori well-defined population such as the treated group. The ATO can be estimated using methods that trim propensity scores, trim treated units, or via overlap weights estimated from propensity scores \citep{Crump:2009,Rosenbaum:2011,li2018balancing}. As such, use of the ATO is typically a response to overlap problems in the data. While analysts might often be more interested in the ATT, choosing the ATO as the estimand allows them at least to estimate \emph{a causal effect} from the data.

In this study, we explore whether a new form of weighting estimator, known as balancing weights, may allow investigators to retain the ATT estimand in applications where overlap is limited. Balancing weights are a generalization of inverse propensity score weights that solve a convex optimization problem to find a set of weights that directly target covariate balance \citep{Hainmueller2011,zubizarreta2015stable}. Recent theoretical work has demonstrated that balancing weights are implicitly estimates of the inverse propensity score, fit via a loss function that guarantees covariate balance \citep{Zhao2016a,Zhao2019,Wang2018,Chattopadhyay2020}. Critically, balancing weights can often outperform inverse propensity score weights estimated from logistic regression \citep{benmichael2022,Chattopadhyay2020,Wang2019}. As such, balancing weights may perform well in many scenarios where IPW fails due to limited overlap.

This paper is organized as follows. In the Methods section, we describe the IPW estimator for the ATT, the overlap weighting (OW) estimator for the ATO, and balancing weights estimator for the ATT. We then conduct three simulation studies to study if balancing weights can recover the ATT in scenarios where IPW may fail. We include comparisons to the OW estimator to understand when balancing weights can recover the same estimand as the ATO. We conclude with a practical application using data on right heart catherization. We show that in this data, balancing weights and OW weights yield the same estimate. Overall, we find that balancing weights expand the set of applications where we can reasonably expect to estimate the ATT, relative to IPW. However, as overlap degrades beyond a certain point, neither can estimate the ATT well. In such settings, it is prudent to switch estimands to the ATO using overlap weights.

\section{Methods}

First, we outline our notation. The population is indexed by units $i=1,\dots,n$, and we denote a binary treatment using $Z_i$ where $Z_{i} = 1$ corresponds to the treated condition and $Z_{i} = 0$ corresponds to the control condition. We let $Y_i$ denote the observed outcome, and use the potential outcomes framework, where each unit has two potential responses: $(Y_{i}(1), Y_{i}(0))$, corresponding to their response under treatment and control, where the observed outcome is $Y_i = Z_iY_i(1) + (1 - Z_i) Y_i(0)$. Next, we assume that the standard assumptions for an observational study based on weighting methods hold. These assumptions are consistency, positivity (overlap), conditional exchangeability, and no interference \citep{Hernan:2020}. Finally, we denote $X_i$ as vector of baseline covariates sufficiently rich to ensure conditional exchangeability holds. 

\subsection{Weighting Estimators of Causal Effects}

Weighting estimators are a popular method for estimating causal effects. For these estimators, each unit is weighted to create a pseudo-population where the baseline covariate distributions between the treated and control units are balanced. This is done through the estimation of weighted averages. Specifically, the estimator is based on a difference in weighted averages across the treated and control groups of the following form: 
\[
\hat{\Delta}_w = \frac{\sum_{i=1}^n Z_iY_iw_i}{Z_iw_i} - \frac{\sum_{i=1}^n (1 - Z_i)Y_iw_i}{(1 - Z_i)w_i},
\]
where we define the weights for each unit as $w_i$. The weights adjust for imbalances in the observed covariates $X_i$.

The weights outlined above can be estimated in two ways. The most common method is the modeling approach. Here, analysts estimate the weights via the propensity score, which is defined as $e(X_i) = P(Z_i = 1|X_i)$ and $\hat{e}(X_i)$ is the estimated propensity score. 
Under the modeling approach, analysts typically estimate the propensity score using logistic regression. That is, the analyst fits the following logistic regression: $1/(1+\exp(-X_i'\beta)$. The coefficients from this model are used to estimate propensity scores for each unit in the study for estimates of $\hat{e}(X_i)$. Once the propensity scores are estimated, the form of the weights depends on the target causal estimand. When the ATE is the causal estimand, the weights are the inverse probability (IP) of receiving treatment or control: $w_i = 1/\hat{e}(X_i)$ for the treated units and $w_i = 1/(1 - \hat{e}(X_i))$ for the control units. When the ATT is the estimand, the weights are $w_i = 1$ for the treated units and the odds of treatment, $w_i = \hat{e}(X_i)/(1 - \hat{e}(X_i))$, for the control units. For the ATO estimand, the weights are $w_i = 1 - \hat{e}(X_i)$ for the treated units and $w_i = \hat{e}(X_i)$ for the control units. 

Under the modeling approach, optimization is based on the fit of the propensity score model, typically through the likelihood. Critically, when the ATE or ATT are the target estimand, the modeling approach to estimating weights suffers from some serious drawbacks. Weights based on logistic regression may not balance covariate distributions very well, and model-based weights can be unstable \citep{Chattopadhyay2020,benmichael2022}. Notably, model based IP weights can be particularly unstable when overlap is poor \citep{Crump:2009,li2018balancing}. On the other hand, model based overlap weights avoid the issues of poor balance and large weights due to poor overlap, but they require focusing on the ATO instead of the ATT \citep{li2018balancing}.

The second approach estimates the weights by targeting the covariate balance in the weighted sample directly. More specifically, estimation of the weights is written as convex optimization problem to find a set of weights that minimizes a combination of a covariate imbalance measure and a penalty term that penalizes large weights. These weights, often referred to as balancing weights, are also based on estimates of the propensity score, but fit via a type of method of moments estimator \citep{benmichael2021_review}. A variety of methods have been proposed for balancing weights including entropy balancing \citep{Hainmueller2011} and stable balancing weights \citep{zubizarreta2015stable} among others \citep{chan2016globally,tan2020regularized,benmichael2021_lor}. Recent research has demonstrated that balancing weights tend to outperform model based weights in both simulations and empirical applications \citep{benmichael2022,Chattopadhyay2020}. One key question is whether improved estimation of the weights via balancing weights can allow analysts to target the ATT in situations where traditional IPW fails. That is, can balancing weights provide an option in some cases to avoid having to change the estimand to the ATO? Next, we study this question via a set of simulation studies.

\section{Simulation}

We conduct a series of simulation studies to investigate how well balancing weights can recover the ATT as overlap varies. In the simulations, we focus on three estimation methods. First, we use IPW estimated via logistic regression. For the model based weights, we use weights for the ATT and the ATO. Next, we estimate balancing weights from \citet{benmichael2022} using the \texttt{balancer} library in R. This method of weighting directly targets covariate imbalance measured as the $L^2$ norm of the weighted difference in means of the covariates, but also includes an $L^2$ regularization term on the sum of the squared weights, which serves as a proxy for the variance of the weighting estimator. These weights include a hyperparameter that controls the bias-variance tradeoff which is set by the user. In many of the simulation scenarios, overlap will be poor.  As such, we set the hyperparameter to prioritize bias reduction in all the simulations. This may inflate the variance of the balancing weights in some scenarios, which would advantage OW.

In the simulation studies, we vary the level of overlap between the treated and control groups and observe when IPW and balancing weights can recover the ATT and when OW recovers the ATO. One complication is that there are a variety of ways to encode overlap into a data generating process (DGP) for the simulations. We do not want our results to be heavily dependent on how overlap is encoded into the DGP. To that end, we use three DGPs from the existing literature that simulate variation in overlap in different ways. For each DGP, we use a sample size of 1000, repeat each simulation scenario 1000 times, and measure bias and root-mean squared error.
In the first two DGPs, the treatment effect will be constant, and so the ATT and the ATO will be equivalent. The final DGP includes treatment effect heterogeneity, considering one case where the ATT and ATO are still equal, and another where they differ.

\subsection{Simulation 1}

In the first simulation, we use the data generating process from \citet{Hainmueller2011} and \citet{Chattopadhyay2020}. In this simulation, there are six observed covariates $X_1, \cdots, X_6$. Covariates $X_1, \cdots, X_3$ are drawn from a mean zero multivariate Normal distribution with the following variance-covariance matrix:
\[
\left[\begin{array}{ccc} 
2 & 1 & -1 \\1 & 1 & -0.5 \\-1 & -0.5 & 1
\end{array}\right]
\]
\noindent Covariates $X_4, \cdots, X_6$ are mutually independent of each other and $X_1, \cdots, X_3$. In the DGP, $X_3 \sim Unif[-3,3]$, $X_5 \sim \chi^2$, and $X_6 \sim Bernoulli(0.5)$. The treatment indicator is generated by the following model: $Z = \mathbbm{1}(X_1 + 2X_2 - 2X_3 - X_4 - 0.5 X_5 + X_6 + \epsilon > 0 )$, where $\mathbbm{1}$ is the indicator function. The error term is drawn from a Normal distribution with mean zero and variance $\sigma^2$.  In this DGP, we can control the degree of overlap through the value of $\sigma^2$, with larger values of $\sigma^2$ leading to better overlap. \citet{Hainmueller2011} and \citet{Chattopadhyay2020} define $\sigma^2 = 100$ as a strong overlap scenario and $\sigma^2 = 30$ as a weak overlap scenario. As such, we vary overlap using the following five values for $\sigma^2$: 20, 40, 60, 80, and 100. Potential outcomes are generated using the following model: $Y(0) = X_1 + X_2 + X_3 -  X_4 + X_5 + X_6 + \eta$, $Y(1) = Y(0) + \delta$, where $\delta$ is the true treatment effect which we set to 5. Here, $\eta$ is drawn from a standard Normal distribution. 

\begin{figure}[h]
  \centering
    \includegraphics[scale=0.7]{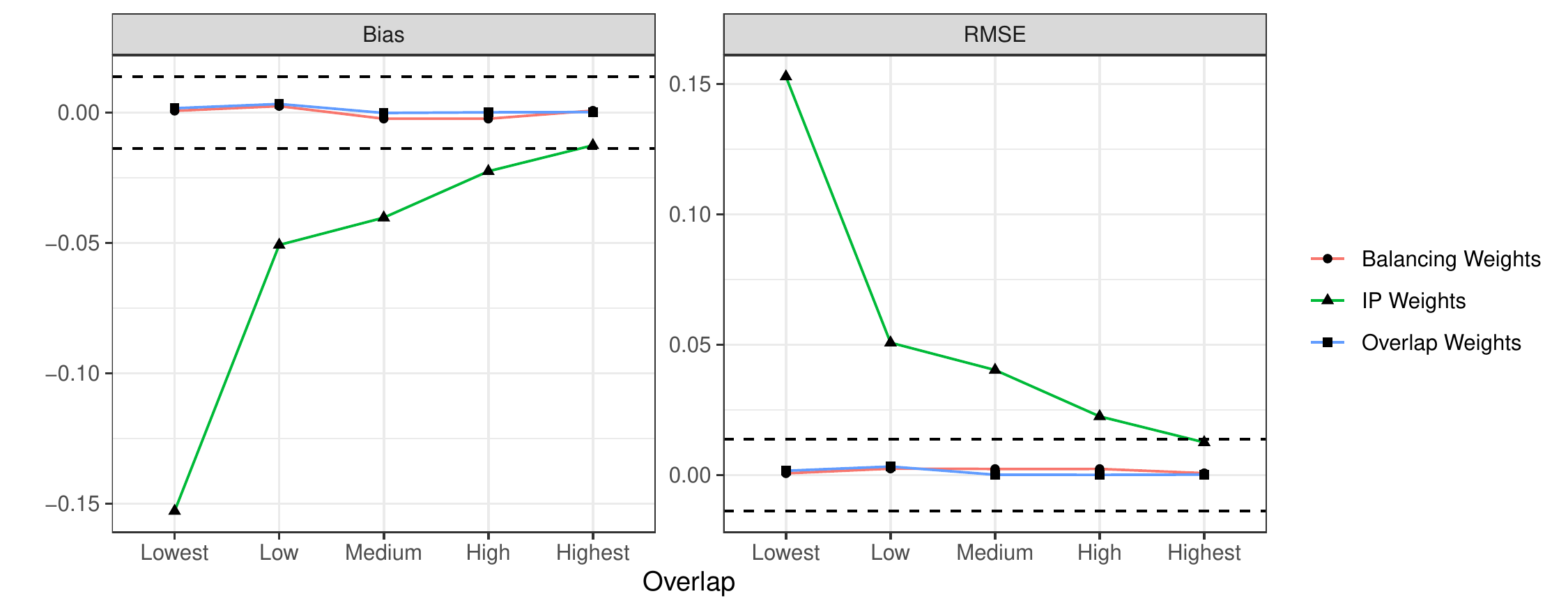}
    \caption{Bias and RMSE for weighting methods in simulation scenario 1. $X$-axis is relative amount of overlap. Higher values of $X$ represent better overlap. Dashed lines represent Monte Carlo error.}
  \label{fig:sim1}
\end{figure}

Figure~\ref{fig:sim1} shows the results from this simulation. We find that balancing weights and overlap weights produce unbiased estimates regardless of the amount of overlap. IP weights on the other hand exhibit large bias when there is little overlap. The amount of bias decreases as overlap improves, but even in the best overlap scenario, the IP weights slightly underestimate the ATT. We also find that both overlap weights and balancing weights have identical performance in terms of RMSE. Here, prioritization of bias reduction does not overly inflate the variance of the balancing weights.

\subsection{Simulation 2}

The next simulation DGP is from \citet{li2019addressing}. For this DGP, we first draw six covariates, $X_1, \cdots, X_6$, from a multivariate normal distribution with zero mean and unit marginal variance, and the correlation between each covariate is 0.5. Covariates $X_1, \cdots, X_3$ are treated as continuous and $X_4, \cdots, X_6$ are converted to binary covariates, so that the marginal mean for each variable is approximately 0.5.  The true propensity score is defined as $e(X) = \{1 + \exp[0.1 + 0.5\gamma X_1 + 0.3\gamma X_2 + 0.3\gamma X_3 - 0.2\gamma X_4 - 0.25\gamma X_5 -0.25\gamma X_6] \}^{-1}$. In this DGP, we use $\gamma$ to control overlap. Larger values of $\gamma$ creates weaker overlap between treated and control groups. The outcome model is defined as a draw from a normal distribution with mean $ -0.5 X_1 - 0.5 X_2 -1.5 X_3 + 0.8 X_4 + 0.8 X_5 1.0 X_6 + \delta Z$ and standard deviation of 2. Following the original simulation, we set $\delta = 0.75$. For $\gamma$, we use values of 2, 3, 4, 5, 6. As before, we record bias and root mean-squared error as performance metrics.

\begin{figure}[h]
  \centering
    \includegraphics[scale=0.7]{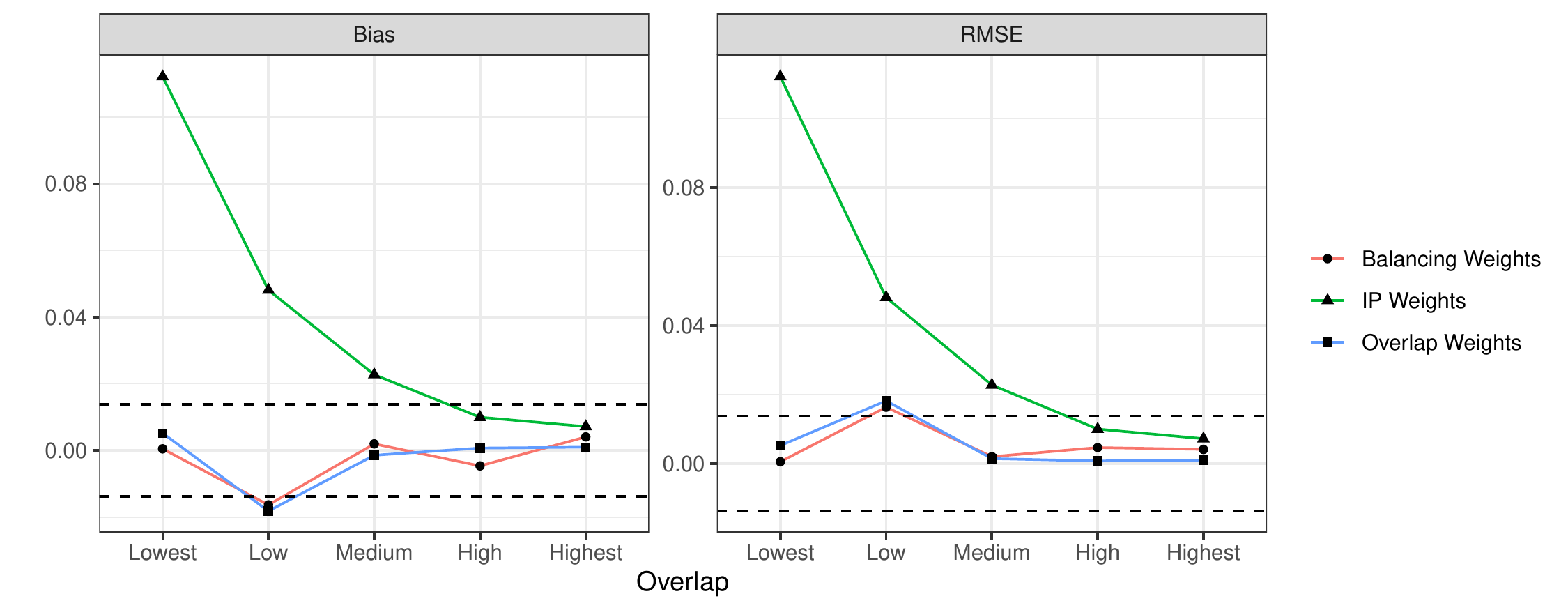}
    \caption{Bias and RMSE for weighting methods simulation scenario 1. $X$-axis is relative amount of overlap. Higher values of $X$ represent worse overlap. Dashed lines represent Monte Carlo error.}
  \label{fig:sim2}
\end{figure}

Figure~\ref{fig:sim2} contains the results from this simulation, which mirror those from simulation 1. We find that balancing weights and overlap weight produce nearly identical results in terms of bias and RMSE. Again, the IP weight results depend strongly on the overlap condition, and estimates based on IP weights tend to be biased unless there is a high level of overlap.

\subsection{Simulation 3}

In the first two simulation DGPs, the treatment effect is constant which implies that the ATT and ATO estimands coincide. As such, any discrepancy thus far has been a function of the estimation method. In the final simulation, we alter the simulation DGP from \citet{benmichael2022} to allow for treatment effect heterogeneity. In this DGP, there are five observed covariates that are independent draws from a multivariate standard Gaussian distribution: $X = (X_1, \dots, X_5) ^\top$. The treatment indicator $Z_i$ is generated from the following model $Z_i = \mathbbm{1}(Z^* > 0)$ where $Z^* = (1.5 X_1 + 1.5 X_2 + .7 X_1 X_2)/c + \text{Unif}(-0.5, 0.5)$. In this DGP, we use $c$ to control the overlap between the treated and control groups. Specifically, we use values of 1, 2.5, 5, 7.5, and 10. When $c=1$ overlap is poor, and when $c=10$, overlap is good. Next, we generate potential outcomes under $Z_i = 0$ as: $Y(0) = X_2 + X_3 + \epsilon_1$ and under $Z_i = 1$ as: $Y(1) = Y(0) + \delta + \delta'$, where $\epsilon_1$ is drawn from a standard Normal distribution and $\delta$ represents a constant treatment effect. We use $\delta'$ to encode non-constant treatment effects. Here, $\delta'$ is a drawn from a normal distribution with expectation $\mu$ and SD of 2. In the simulations, we set $\delta= 5$ and vary the model for $\mu$. In the first simulation, we set $\mu = .5 X_4 + .25 X_5$.  Here, the treatment effect heterogeneity is a function of two covariates that don't enter the propensity score model. For this DGP, while the treatment effects are non-constant, the ATT and ATO will be equivalent. As such, differences across estimators will be a function of how well the weights are estimated. In the second simulation, we specify $\mu = .5 X_1 + .25 X_2$. Now, the heterogeneity is a function of the two covariates that enter the propensity score model. For this DGP, the ATT and ATO will differ regardless of the amount of overlap. In the simulations, we calculate the true ATT and true ATO for each simulation run, and measure bias relative to the appropriate true value. Here, we also plot the average estimate for each method as well as bias, so the reader can see the divergence between the two estimands.

\begin{figure}[h]
  \centering
    \includegraphics[scale=0.7]{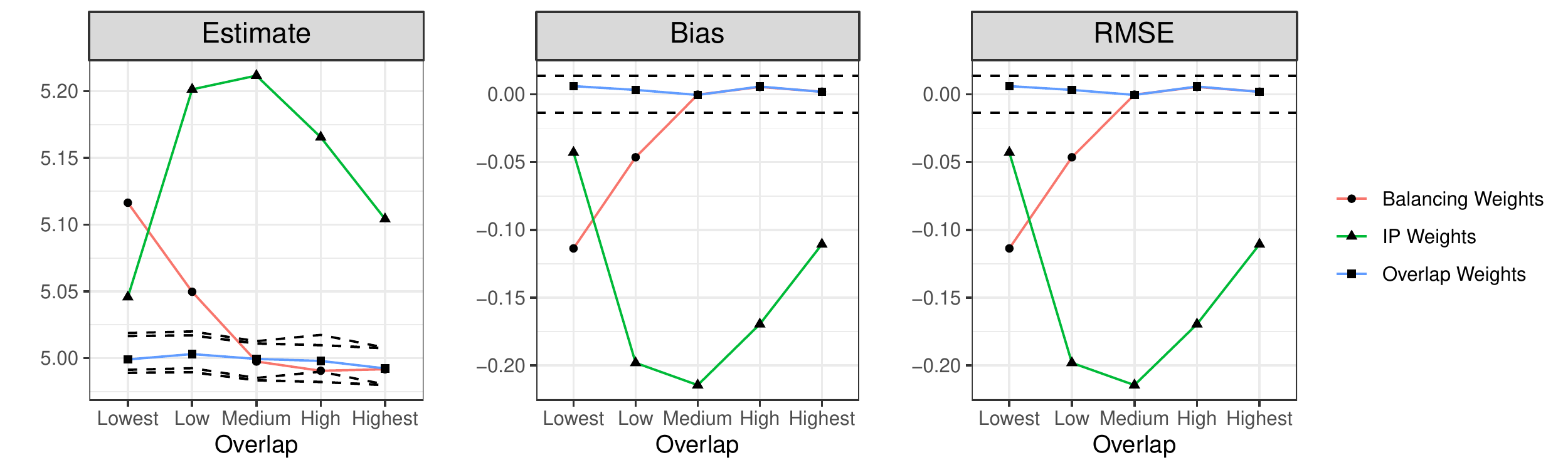}
    \caption{Bias and RMSE for weighting methods simulation scenario 3: overlap unrelated to propensity score. $X$-axis is relative amount of overlap. Higher values of $X$ represent worse overlap. Dashed lines represent Monte Carlo error.}
  \label{fig:sim3-1}
\end{figure}

Figure~\ref{fig:sim3-1} contains the results from the first scenario, where the ATT and ATO are equal. Here, we find that balancing weights display some levels of bias for the two lowest levels of overlap. Estimates based on IP weights, however, are biased across all levels of overlap.
So we see that balancing weights can estimate the ATT well in settings where traditional IPW cannot, but when overlap is poor, using overlap weights and targeting the ATO estimand allows us to recover a valid causal effect.

\begin{figure}[h]
  \centering
    \includegraphics[scale=0.7]{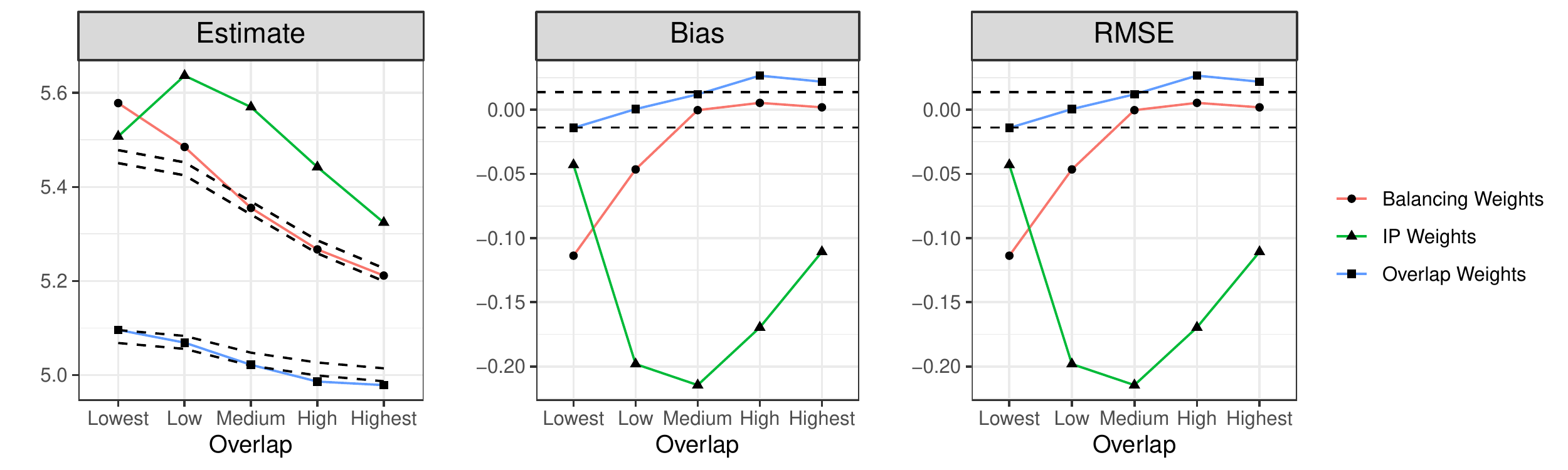}
    \caption{Bias and RMSE for weighting methods simulation scenario 3: overlap related to propensity score. $X$-axis is relative amount of overlap. Higher values of $X$ represent worse overlap. Dashed lines represent Monte Carlo error.}
  \label{fig:sim3-2}
\end{figure}

Figure~\ref{fig:sim3-2} contains the results from the second scenario, where treatment effects correlate with the propensity scores. As expected, here the ATO and ATT no longer correspond. 
In this setting we find similar results as in Figure~\ref{fig:sim3-1} above: traditional IPW estimates the ATT poorly, while balancing weights yield good estimates when overlap is moderate to good, with the performance degrading in the low overlap settings.
And again, overlap weights perform well across all levels of overlap.
Note however, that while switching to the ATO in low overlap settings will lead to a substantial difference in the estimand, and so care must be given to the changing interpretation of the estimand, unlike in the simulation settings above.

\section{Application}

Next, we conduct an empirical analysis examining the effectiveness of right heart catherization (RHC), a monitoring device that is used in the management of critically ill patients. While RHC can provide important diagnostic information, it may also contribute to the onset of complications or death. The SUPPORT study gathered data on the use of RHC at five medical centers in the U.S. on 5735 critically ill patients that were admitted to the intensive care unit (ICU) \citep{murphy1990}. In the data, 2184 patients received RHC within 24 hours of admission, the rest are treated as control patients. The primary outcome in the data is mortality at 30 days after admission to the ICU. The data include a large number of covariates that a set of clinical experts selected as having prognostic value. One early innovative study used this data with propensity score matching \citep{connors1996effectiveness}. Overlap in these data is known to be poor, and these data have been reanalyzed numerous times to develop estimation methods for treatment effects in such contexts \citep{Crump:2009,traskin2011defining,li2018balancing}.

In our re-analysis, we compare treatment effect estimates based on three different methods. First, we estimate the propensity score using a logistic regression, using these propensity score estimates to calculate ATT and overlap weights. We then calculate ATT weights using  balancing weights. For each method, we calculate a metric of bias reduction called percentage of bias reduction (PBR), which measures how much weighting reduces imbalance compared to the unweighted data. We calculate PBR using standardized differences. For covariate $k$, the standardized difference before weighting is
\[
\hat{\Delta}_{k} = \frac{\bar{X}_{tk} - \bar{X}_{ck}}{\sqrt{(V(X_{tk}) + V(X_{tc}))/2}}
\]
\noindent where $X_{tk}$ and $X_{ck}$ are the treated and control group vectors for covariate $k$. The standardized difference after weighting is
\[
\hat{\Delta}_{wk} = \frac{\bar{X}_{wtk} - \bar{X}_{wck}}{\sqrt{(V(X_{tk}) + V(X_{tc}))/2}}
\]
\noindent where $\bar{X}_{wtk}$ and $\bar{X}_{wck}$ are weighted means based on the estimated weights. We refer to the vector of standardized differences for all $K$ covariates in the unweighted data as $\hat{\Delta}_{uw}$, and we use $\hat{\Delta}_{w}$ to denote the vector of standardized differences for all $K$ covariates in the weighted data. We measure PBR as:
\[ 
PBR = 100\% \times \left[ \frac{1}{K} \sum_k |\hat{\Delta}_{w}| \; \Big/ \; \frac{1}{K} \sum_k |\hat{\Delta}_{uw}| \right]  .
\]
\noindent This measure describes the reduction in bias based on the change in balance across all covariates due to weighting.

Table~\ref{tab.rhc} contains the estimated treatment effect for each method on a risk difference scale, the estimated standard error, and the PBR. First, we note that balancing weights are have the highest PBR at 100\%.  This provides further evidence on how well balancing weights can reduce bias. IPW, on the other hand, produces a PBR of 81\%. Next, we find that balancing weights and overlap weights produce nearly identical point estimates. The key advantage of the balancing weights is in terms of interpretation. Here, the ATT represents the effect of RHC on those patients that actually received RHC. The estimate based on the ATO is the effect of RHC for those treated patients that were marginal for treatment via RHC. As such, by altering the estimation method of the weights, we are able to retain the interpretability of the ATT estimand.

\begin{table}[ht]
\centering
\caption{Results for the RHC Application Using Three Different Estimation Methods}
\label{tab.rhc}
\begin{tabular}{lccc}
  \toprule
& Bal. Weights - ATT & Overlap Weights & IPW- ATT \\ 
  \midrule
Point Estimate  & 0.065 & 0.067 & 0.062  \\ 
Standard Error  &  0.016 & 0.015 & 0.021  \\ 
Bias Reduction (\%)  &  100 & 96.4 & 81.3  \\ 
   \bottomrule
\end{tabular}
\end{table}

\section{Discussion}

In this study, we explored whether balancing weights allow analysts to target the ATT estimand more often in situations where the overlap between the treated and control covariate distributions is limited. It is well known that standard IPW methods can be biased with larger variances in such situations. OW weighting methods are one solution when overlap is poor, but it requires targeting an estimand that may be harder to interpret. We found that balancing weights, by directly targeting imbalances in the covariate distributions, produced lower bias even as overlap was decreasing. This reduction in bias often makes the ATT a feasible target estimand even when overlap is not high. Moreover, we found that balancing weights outperformed IPW methods in almost every situation. In many cases, IPW estimates were biased, while balancing weights were not. Our study provides further evidence that balancing weights generally outperform IPW and should generally be preferred in applied analyses. Balancing weights offer a superior alternative in many situations where overlap is less than perfect. In our simulations, IPW methods only performed well when overlap was at its highest level.

Our results also demonstrate that in some cases, particularly when treatment effects were non-constant, it is necessary to use OW and target the ATO estimand. This suggests that balancing weights and OW are complementary strategies for observational studies.  Balancing weights expand the range of applications where the ATT is applicable. However, in applications with poor overlap or when effects are nonconstant, OW allow analysts to target the ATO which is still identifiable when the ATT is not.

Our study has some limitations. We only focused on settings without longitudinal variation in the treatment. We also only focused on the setting where the conditional exchangeability holds. We did not consider the case where additional outcome modeling may be necessary to reduce bias. We also did not consider how specification of the basis function -- the functional form of the baseline covariate matrix -- might affect the results. Other research has demonstrated that the performance of balancing weights will be improved when higher moments and nonlinearities are included in the basis function \citep{huang2022higher,setodji2017right}. \citet{benmichael2022} demonstrate how to fully specify the basis function to capture both higher moments and possible interactions between covariates using machine learning methods. Full specification of the basis function can be critical to eliminating bias from model misspecification.

Our results suggest the following work flow for applied analysts. First, investigators should determine which target estimand is preferred given the scientific question of interest. Next, tables of baseline characteristics and summary plots of distributions can be used to diagnose the extent to which treated and control distributions overlap. When there is a high amount of overlap between treated and control distributions, empirical considerations need not influence the choice of the estimand. However, when overlap is appears to be lacking, investigators can conduct two parallel analyses: one based on balancing weights targeting the ATT and one based on OW weights targeting the ATO. When results from the two analyses agree, this provides evidence that the ATT estimand is identifiable. When the two estimates diverge, then the ATO should be considered the identifiable estimand. In both cases, analysts should be sure to include a table that describes the weighted pseudo population compared to the unweighted population. This is particularly critical when the ATO is the estimand, since the treated group can no longer be defined a priori.

\clearpage
\renewcommand{\refname}{References Cited}

\singlespacing
\bibliography{ob-weight}

\end{document}